\documentclass[12pt]{article}
\usepackage{epsfig,graphicx}
\usepackage{fpcp03}

\def\beq{\begin{equation}}
\def\eeq#1{\label{#1}\end{equation}}
\def\eeqn{\end{equation}}

\def\beqa{\begin{eqnarray}}
\def\eeqa#1{\label{#1}\end{eqnarray}}
\def\eeqan{\end{eqnarray}}

\let\bar=\overbar

\def\Dslash{\not{\hbox{\kern-4pt $D$}}}
\def\dslash{\not{\hbox{\kern-2pt $\del$}}}

\def\msb{{\bar{\ssstyle M \kern -1pt S}}}

\def\BB0bar{B^0 {\overline B}^0}
\def\BB0dbar{B_d^0 {\overline B}_d^0}
\def\BB0sbar{B_s^0 {\overline B}_s^0}

\RequirePackage{xspace}

\usepackage{relsize}
\def\babar{\mbox{\slshape B\kern-0.1em{\smaller A}\kern-0.1em
    B\kern-0.1em{\smaller A\kern-0.2em R}}}

\def\Kbar  {\kern 0.2em\overline{\kern -0.2em K}{}\xspace}

\def\Kz    {\ensuremath{K^0}\xspace}
\def\Kzb   {\ensuremath{\Kbar^0}\xspace}
\def\KzKzb {\ensuremath{\Kz \kern -0.16em \Kzb}\xspace}
\def\Kp    {\ensuremath{K^+}\xspace}
\def\Km    {\ensuremath{K^-}\xspace}

\def\KpKm  {\ensuremath{\Kp \kern -0.16em \Km}\xspace}

\def\Dbar    {\kern 0.2em\overline{\kern -0.2em D}{}\xspace}

\def\Dz      {\ensuremath{D^0}\xspace}
\def\Dzb     {\ensuremath{\Dbar^0}\xspace}
\def\DzDzb   {\ensuremath{\Dz {\kern -0.16em \Dzb}}\xspace}
\def\Dp      {\ensuremath{D^+}\xspace}
\def\Dm      {\ensuremath{D^-}\xspace}

\def\DpDm    {\ensuremath{\Dp {\kern -0.16em \Dm}}\xspace}

\def\Bbar    {\kern 0.18em\overline{\kern -0.18em B}{}\xspace}

\def\BB      {\ensuremath{B\Bbar}\xspace} 
\def\Bz      {\ensuremath{B^0}\xspace}
\def\Bzb     {\ensuremath{\Bbar^0}\xspace}
\def\BzBzb   {\ensuremath{\Bz {\kern -0.16em \Bzb}}\xspace}
\def\Bu      {\ensuremath{B^+}\xspace}
\def\Bub     {\ensuremath{B^-}\xspace}

\def\BpBm    {\ensuremath{\Bu {\kern -0.16em \Bub}}\xspace}

\mathchardef\Upsilon="7107
\def\Y#1S{\ensuremath{\Upsilon{(#1S)}}\xspace}

\mathchardef\Deltares="7101
\mathchardef\Xi="7104
\mathchardef\Lambda="7103
\mathchardef\Sigma="7106
\mathchardef\Omega="710A

\def\Deltabar{\kern 0.25em\overline{\kern -0.25em \Deltares}{}\xspace}
\def\Lbar{\kern 0.2em\overline{\kern -0.2em\Lambda\kern 0.05em}\kern-0.05em{}\xspace}
\def\Sigbar{\kern 0.2em\overline{\kern -0.2em \Sigma}{}\xspace}
\def\Xibar{\kern 0.2em\overline{\kern -0.2em \Xi}{}\xspace}
\def\Obar{\kern 0.2em\overline{\kern -0.2em \Omega}{}\xspace}
\def\Nbar{\kern 0.2em\overline{\kern -0.2em N}{}\xspace}
\def\Xb{\kern 0.2em\overline{\kern -0.2em X}{}\xspace}

\newcommand{\tev}{\ensuremath{\mathrm{\,Te\kern -0.1em V}}\xspace}
\newcommand{\gev}{\ensuremath{\mathrm{\,Ge\kern -0.1em V}}\xspace}
\newcommand{\mev}{\ensuremath{\mathrm{\,Me\kern -0.1em V}}\xspace}
\newcommand{\kev}{\ensuremath{\mathrm{\,ke\kern -0.1em V}}\xspace}
\newcommand{\ev}{\ensuremath{\mathrm{\,e\kern -0.1em V}}\xspace}
\newcommand{\gevc}{\ensuremath{{\mathrm{\,Ge\kern -0.1em V\!/}c}}\xspace}
\newcommand{\mevc}{\ensuremath{{\mathrm{\,Me\kern -0.1em V\!/}c}}\xspace}
\newcommand{\gevcc}{\ensuremath{{\mathrm{\,Ge\kern -0.1em V\!/}c^2}}\xspace}
\newcommand{\mevcc}{\ensuremath{{\mathrm{\,Me\kern -0.1em V\!/}c^2}}\xspace}

\def\mus  {\ensuremath{\rm \,\mus}\xspace}

\def\mus        {\ensuremath{\,\mu{\rm s}}\xspace}    

\def\to                 {\ensuremath{\rightarrow}\xspace}

\def\pep2{PEP-II}

\def\gsim{{~\raise.15em\hbox{$>$}\kern-.85em
          \lower.35em\hbox{$\sim$}~}\xspace}
\def\lsim{{~\raise.15em\hbox{$<$}\kern-.85em
          \lower.35em\hbox{$\sim$}~}\xspace}

\xspace

\def\jetset74   {\mbox{\tt Jetset \hspace{-0.5em}7.\hspace{-0.2em}4}\xspace}

\def\Dsjzero     {\ensuremath{D^{*}_{sJ}(2317)}\xspace}
\def\Dsjone      {\ensuremath{D_{sJ}(2463)}\xspace}
\begin{document}


\Title{\boldmath Observation of the \Dsjone and Confirmation of the
                 \Dsjzero }
\bigskip


%
\label{StoneStart}

\author{Sheldon Stone$^{\it a}$ and Jon Urheim$^{\it b}$\index{Stone, S.} }

%
\address{Physics Department$^{\it a}$ \\
Syracuse, University,\\
Syracuse, NY 13244-1130, USA\\
Physics Department$^{\it b}$ \\
University of Minnesota\\
Minneapolis, MN 55455, USA}

\makeauthor\abstracts{Using 13.5 fb$^{-1}$ of $e^+e^-$
annihilation data in the CLEO II detector at CESR, we have
observed a new narrow state decaying to $D_s^{*+}\pi^o$, denoted
the $D_{sJ}(2463)^+$.  A possible interpretation holds that this
is a $J^P = 1^+$ partner to the $D_{sJ}^*(2317)^+$ state recently
discovered by the BaBar Collaboration which is consistent with
$J^P = 0^+$. We have also confirmed the existence of the
$D_{sJ}^*(2317)^+$ in its decay to $D_s^+\pi^o$.  We have measured
the masses of both states, accounting for the cross-feed
background that the two states represent for each other, and have
searched for other decay channels for both states. No narrow
resonances are seen in $D_s^{\pm}\pi^{\mp}$ or
$D_s^{\pm}\pi^{\pm}$ modes.}

\section{Introduction}

Prior to this year, the spectrum of $c\overline{s}$ mesons
was believed to be well-understood.
The weakly-decaying ground state $D_s^+$ meson with mass
1969 MeV and $J^P = 0^-$ was discovered by CLEO in 1983.
The excited $1^-$ state at 2112.4 MeV, the $D_s^{*+}$ meson,
is also narrow, decaying to the $D_s^+$ predominately via $\gamma$
emission.  It also has a $6\%$ rate~\cite{CLEO_dspi0} for a strong
transition via $\pi^o$ emission~\cite{Cho-Wise},
which violates isospin symmetry since all $c\overline{s}$ mesons
are isospin singlets while the pion is an isospin triplet.
These states both have zero orbital angular momentum
between the two quarks.

Four states with $L=1$ are expected, corresponding to a spin singlet
and triplet, giving one state with $J^P = 0^+$, two with $1^+$,
and one with $2^+$.  Considering the charm quark to be heavy,
it is more natural to think of these as two doublets with $j=1/2$ and 3/2,
where $j$ is the angular momentum sum of $L$ with the spin of the strange
quark.  The $j=3/2$ states are expected to be narrow because
their dominant (OZI- and isospin-favored) decays to $D^{(*)}K$ will
proceed via $D$-wave.  Indeed, the experimental observations of the
$D_{sJ}(2573)^+$ (with $J^P$ consistent with $2^+$) and the $J^P=1^+$
$D_{s1}(2536)^+$ were made feasible by the fact that these
states are narrow.
Most, but not all, potential models expected the unobserved $j=1/2$
states to have comparable masses, and to decay to same final states
but with large widths, $\sim$200-300 MeV, since these decays would
proceed via $S$-wave.

BaBar has recently reported the discovery of a new narrow state,
the $D_{sJ}^*(2317)^+$, in its decay to
$D_s^+\pi^o$~\cite{BaBarDss}, its width consistent with
experimental resolution. The low mass, below $DK$ threshold,
implies that despite its isospin violation, the observed channel
is the most likely hadronic decay available, thus explaining the
narrow width. The BaBar data is also consistent with a $0^+$
spin/parity interpretation.

Various interpretations of this state have appeared in the
literature. To give some examples: Barnes, Close and Lipkin
speculate that this could be ``baryonia" or a $DK$
molecule~\cite{Barnes}. Van Beveran and Rupp suggest a quasi bound
scalar that arises due to coupling to the nearby $DK$
threshold~\cite{VB}. Cahn and Jackson formulate an acknowledgely
poor explanation using non-relativistic vector and scalar exchange
forces~\cite{Cahn}.

Bardeen, Eichten and Hill (BEH)~\cite{bardeeneichtenhill} use HQET plus
chiral symmetry to predict ``parity doubling," where two orthogonal linear
combinations of mesons transform as SU(3)$_L\times$SU(3)$_R$ and split
into ($0^-,1^-$), ($0^+,1^+$) doublets.  Assuming that the $D_{sJ}^*(2317)$
is the $0^+$ state expected in the quark model, their concrete prediction
is that the mass splitting between the remaining $1^+$ state and the $1^-$
should be the same as the $0^+-0^-$ splitting.

\section{\boldmath Confirmation of the \Dsjzero}
\begin{figure}[htb]
\begin{center}
  \epsfig{file=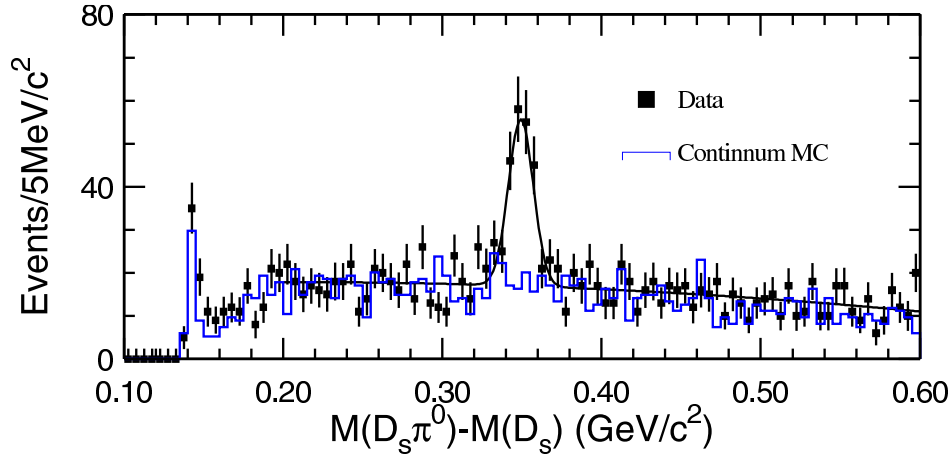,height=60mm}
  \caption{The $D_s^+\pi^o$ candidate mass distribution
           shown as the difference with respect to the $D_s^+$ mass.
           The curve shows our Monte Carlo simulation of the
           spectrum, absolutely normalized,
           without narrow states decaying into $D_s^+$.}
  \label{Dspi0_new-1}
\end{center}
\end{figure}

$D_s^+$ candidates are looked for in the $\phi\pi^+$ decay mode.
The selection criteria are described in detail in Ref.~\cite{CLEO_Dss}.
The $D_s^+\pi^o$ mass distribution is shown in Fig.~\ref{Dspi0_new-1}
for mass combinations with momenta above 3.5 GeV/c.
Two peaks are evident: one near a mass difference of 0.1 GeV,
due to the decay of the $D_s^{*+}$ into a $D_s^+\pi^o$, and another,
larger structure near a mass difference of 0.35 GeV, that confirms the
existence of the $D_{sJ}^*(2317)^+$. The measured width of this peak
is $8.0^{+1.3}_{-1.2}$ MeV, somewhat wider than the detector resolution
of 6.0$\pm$0.3 MeV.  The curve shows our Monte Carlo simulation of the
mass distribution, absolutely normalized, without the presence of any
narrow states that decay into $D_s^+$ mesons.  The CLEO Monte Carlo
does an excellent job of reproducing the size and shape of our background
$\pi^o$ candidates.

The peak near 0.35 GeV is close to the mass reported by BaBar. Our mass
determination will be discussed later.  We observe 165$\pm$20 events in
this peak.

\section{\boldmath Observation of the \Dsjone}
\begin{figure}[htb]
\begin{center}
\epsfig{file=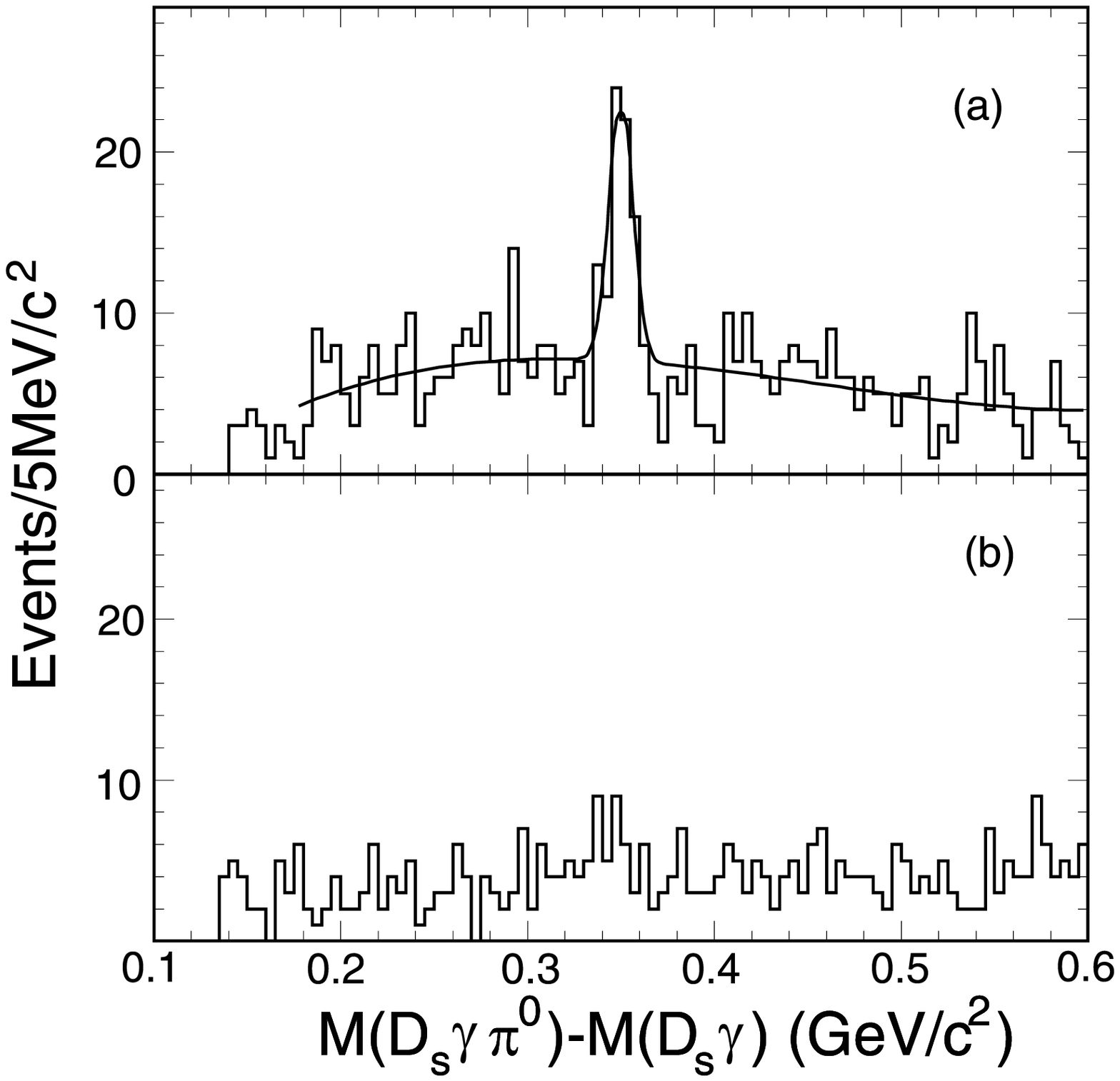,height=80mm}
\caption{The $D_s^{*+}\pi^o$ candidate mass distribution shown
as the difference with respect to the $D_s^{*+}$ mass.
(a) $D_s^{*+}$ signal region; (b) $D_s^{*+}$ sideband region. }
\label{data_dsspi0_2}
\end{center}
\end{figure}

We also looked for decays of the $D_{sJ}^*(2317)^+$
and possible additional narrow states in other
channels, notably $D_s^{*+}\pi^o$.
We use the $D^{*+}\to\gamma D_s^+$ decay mode.
Photon candidates were selected from neutral energy clusters
with lateral profiles consistent with electromagnetic showers
and absolute energies above 50 MeV.  Fig.~\ref{data_dsspi0_2}
shows the mass difference distributions for both the peak and
sideband regions of the $D_s^{*+}$ signal.

We observe a peak consisting of 55$\pm$10 events,
with a width of 6$\pm$1.0 MeV (r.m.s.) compared with the detector
resolution of 6.6$\pm$0.5 MeV.
The mass difference value is also about 0.35 GeV.  The near equality
of this mass difference with the previous one leads to the worry
that there could be cross-contamination between the two final states.

\section{Analysis of Cross Contamination}

Many studies were performed to see if these two states could arise
from reflections of other known narrow states.  These possibilities
were excluded.

It is possible, however, for a higher mass state decaying $D_s^{*+}\pi^o$
to be reconstructed as a lower mass state simply by ignoring the photon
from the
$D_s^{*+}$ decay.  In fact, taking the signal $D_s^{*+}\pi^o$ events
and ignoring the photon from the $D_s^{*+}$ decay causes a peak in
the $D_s^{+}\pi^o$ spectrum at very nearly the same mass difference,
but with a width of 14.9 MeV, considerably larger than our resolution.
The efficiency of this process is rather high: (84$\pm$4$\pm$10)\%.

It is also possible for the lower mass state to pick up a random
photon, fake a $D_s^{*+}$, and thus be a candidate for the upper
mass state. This is a much smaller probability,
(9.0$\pm$0.7$\pm$1.5)\% and can be estimated from the $D_s^{*+}$
sidebands.  The number of actual signal events can be estimated
from these probabilities and the measured numbers of events in the
peaks.  Accounting for the background in this way, the peak in the
$D_s^{*+}\pi^o$ sample corresponds to 41$\pm$12 signal events. The
probability that this excess is due to a background fluctuation is
in excess of 5$\sigma$. Thus CLEO has made the first observation
of a new state near 2460 MeV.  (Although the BaBar data also
showed an excess of events in this mass region, the conclusion
reached in Ref.~\cite{BaBarDss} was that further study was needed
to resolve whether the peak received contributions from a new
state or was entirely due to a reflection of the
$D_{sJ}^*(2317)$.)

\section{Mass Determinations}

Because of the contamination of the lower mass state by the
higher mass one, fitting the $D_s\pi^o - D_s$ mass difference
distribution to a single Gaussian could result in a biased mass
determination.  Taking advantage of the the excellent
mass resolution of the CLEO CsI calorimeter we fit the $D_s^+\pi^o$
mass difference peak to two Gaussians whose means and widths
are allowed to float.  The fit determines one signal to be at a
mean mass difference of 350.0$\pm$1.2 MeV with a width of
5.9$\pm$1.2 MeV and another wider Gaussian at 344.9$\pm$6.1 MeV
with a width of 16.5$\pm$6.3 MeV, characteristic of the feed-down
background.  We use this fit for our determination of the mass
difference, to which we assign a $\pm$1.0 MeV systematic error.

\begin{figure}[htb]
\begin{center}
\epsfig{file= 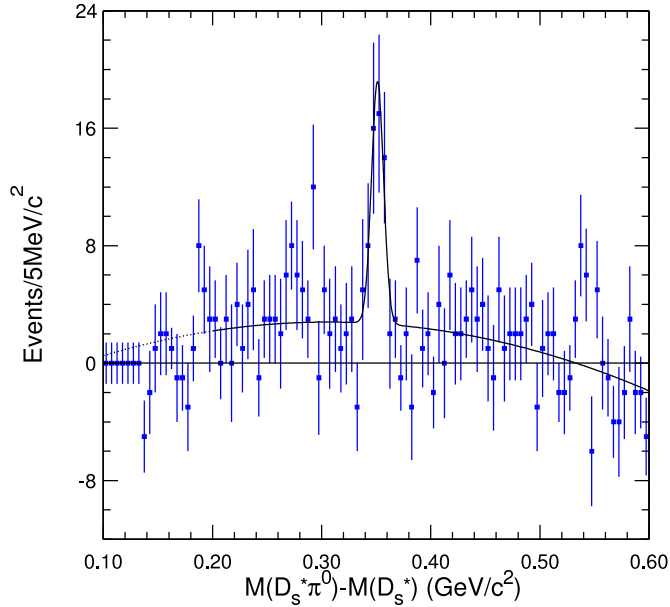,height=80mm}
\caption{The sideband subtracted $D_s^{*+}\pi^o$ candidate mass
difference distribution. The curve resents a fit to a signal
Gaussian whose mean and width are allowed to float and a second
order background polynomial.}
\label{dsstpio_data_fit_dm_ss2}
\end{center}
\end{figure}

Since the feedup from the first state to the second state is
relatively small, $\sim$20\% of the signal of the higher mass
state, we determine its mass by subtracting the $D_s^*$ sidebands and
performing a fit.  The subtracted spectrum and the fit are shown in
Fig.~\ref{dsstpio_data_fit_dm_ss2}.  The resulting mass difference
is 351.2$\pm$1.7 MeV, to which we also assign a systematic error
of $\pm$1.0 MeV.

We note that a $D_s^+\pi^o$ system with $L=0$ is a $0^+$
state, and a $D_s^{*+}\pi^o$ system with $L=0$ is a $1^+$ state. If
the \Dsjone were a $0^+$ state it would be above threshold for
decay into $DK$.  There is no evidence for this state in that decay
mode and if that decay occurred the state would be wide.

\section{Upper Limits On Other Decay Modes}

\subsection{Neutral and Doubly Charged Modes}

\begin{figure}[htb]
\begin{center}
  \epsfig{file=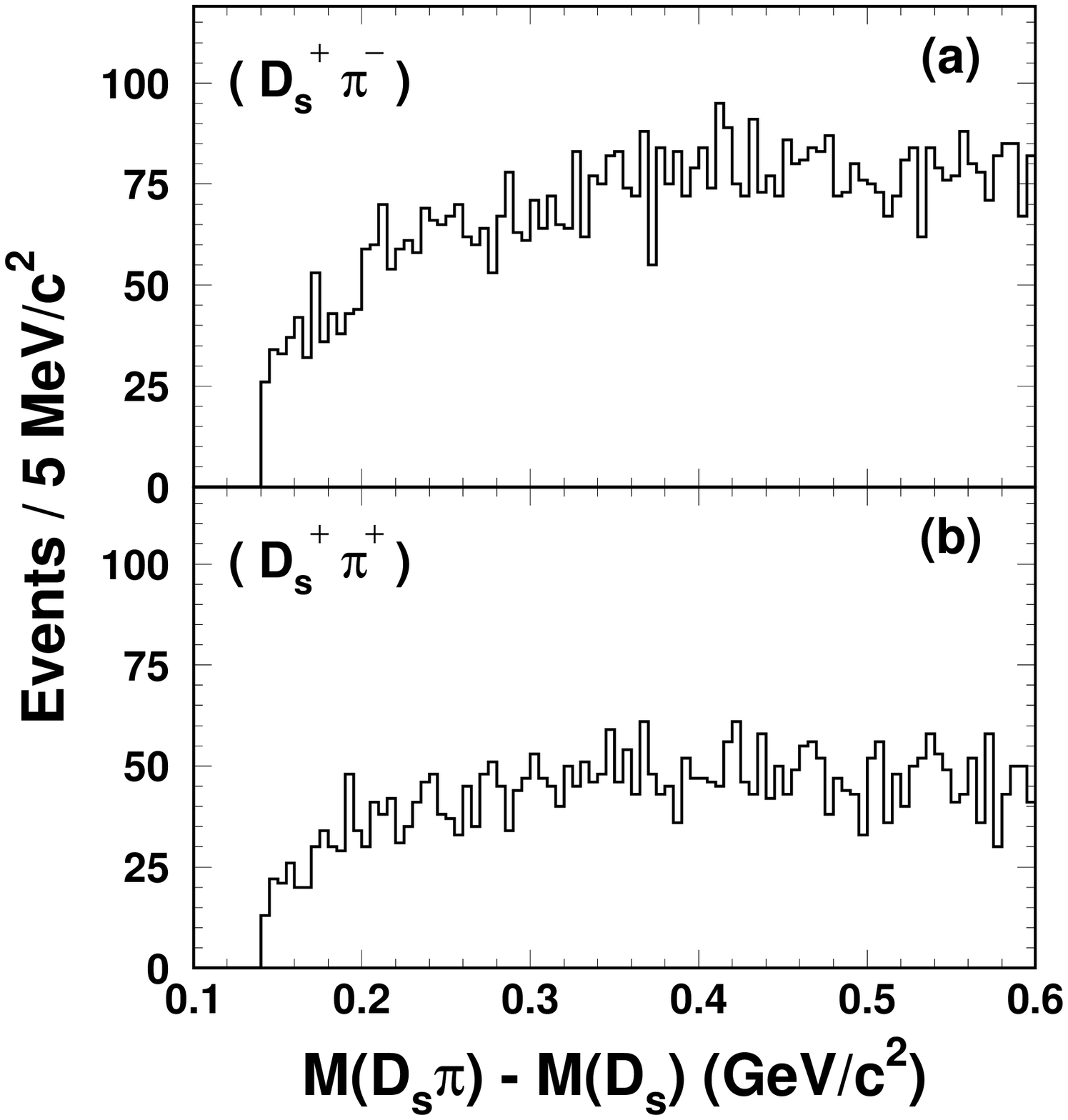,height=80mm}
  \caption{Mass difference distributions for $D_s^\pm\pi^\mp$ (top)
           and $D_s^\pm\pi^\pm$ (bottom) candidate samples.
          }
  \label{fig:dspic}
\end{center}
\end{figure}

In Fig.~\ref{fig:dspic} we show the $D_s^{\pm}\pi^{\mp}$ and
$D_s^{\pm}\pi^{\pm}$ mass difference distributions.
No signals are visible and the production ratio times decay rate
of any objects similar to the $D_{sJ}^*(2317)$ are lower by more
than a factor of ten compared to the $D_s^{*+}\pi^o$ mode.
This argues against a molecular interpretation.

\subsection{\boldmath Other Decay Modes of the \Dsjzero}

Upper limits on other decay modes relative to the $D_s^+\pi^o$
mode are given in Table~\ref{tab:limitszero}.
\begin{table}
\begin{center}
  \begin{tabular}{|lrccc|}
  \hline
  Final State & Yield & Efficiency & Ratio ($90\%$ C.L.) & Prediction \\
  \hline
  $D_s^+ \pi^0 $           & $135\pm 23$   & $(9.7 \pm 0.6)\,\%$ &  ---
                           & \\
  $D_s^+ \gamma$           & $-19\pm 13$   & $(18.1 \pm 0.1)\,\%$ & $< 0.052 $
                           & 0 \\
  $D_s^{*+} \gamma$        & $-6.5\pm 5.2$ & $( 7.0 \pm 0.5)\,\%$ & $< 0.059 $
                           & 0.08 \\
  $D_s^+ \pi^+\pi^-$       & $2.0\pm 2.3$  & $(19.8 \pm 0.8)\,\%$ & $< 0.019 $
                           & 0 \\
  $D_s^{*+} \pi^0$         & $-1.7\pm 3.9$ & $( 3.6 \pm 0.3)\,\%$ & $< 0.11 $
                           & 0 \\
  \hline
  \end{tabular}
  \caption{The 90\% C.L.\ upper limits on the ratio
  of branching fractions for $D_{sJ}^*(2317)$ to the the channels
  shown relative to the $D_s^+\pi^0$ state.  Also shown are the
  theoretical expectations from Ref.~\cite{bardeeneichtenhill},
  under the assumption that the $D_{sJ}^*(2317)$ is the
  lowest-lying $0^+$ $c\overline{s}$ meson.}
  \label{tab:limitszero}
\end{center}
\end{table}

\subsection{\boldmath Other Decay Modes of the \Dsjone}

Limits obtained on other decays, relative to $D_s^*\pi^0$, are
summarized in Table~\ref{tab:limitsone}.
\begin{table}
\begin{center}
  \begin{tabular}{|lrccc|}
  \hline
  Final State & Yield & Efficiency & Ratio ($90\%$ C.L.) & Prediction \\
  \hline
  $D_s^{*+} \pi^0$         & $41\pm 12$    & $(6.0 \pm 0.2)\,\%$ &  ---
                           & \\
  $D_s^+ \gamma$           & $ 40\pm 17$   & $(19.8 \pm 0.4)\,\%$ & $< 0.49 $
                           & 0.24 \\
  $D_s^{*+} \gamma$        & $-5.1\pm 7.7$ & $( 9.1 \pm 0.3)\,\%$ & $< 0.16 $
                           & 0.22 \\
  $D_s^+ \pi^+\pi^-$       & $2.5\pm 5.4$  & $(19.5 \pm 1.5)\,\%$ & $< 0.08 $
                           & 0.20 \\
  $D_{sJ}^*(2317)^+\gamma$ & $3.6\pm 3.0$  & $(2.0 \pm 0.1)\,\%$  & $< 0.58 $
                           & 0.13 \\
  \hline
  \end{tabular}
  \caption{The 90\% C.L.\ upper limits on the ratio
  of branching fractions for $D_{sJ}(2463)$ to the the channels
  shown relative to the $D_s^{*+}\pi^0$ state.  Also shown are the
  theoretical expectations from Ref.~\cite{bardeeneichtenhill},
  under the assumption that the $D_{sJ}(2463)$ is the
  lowest-lying $1^+$ $c\overline{s}$ meson.}
  \label{tab:limitsone}
\end{center}
\end{table}

This electromagnetic transition to
$D_{sJ}^*(2317)^+\gamma$~\cite{noteonxfeed} presents a
particularly difficult situation as the final state particles are
again a $D_s^+$ a $\pi^o$ and a $\gamma$ with momenta similar to
that in the main $D_s^{*+}\pi^o$ mode. To reduce backgrounds from
$D_{sJ}(2463)^+\to D_s^{*+}\gamma$, we required that the
$D_s\pi^0$ system be consistent with the decay of the
$D_{sJ}^*(2317)$, namely that $|\Delta M(D_s\pi^0) -
350.0\,\mbox{\rm MeV/c}^2| < 13.4\,$MeV/c$^2$ ($\sim 2\sigma$
based on Monte Carlo simulations). We also required that the
$D_s\gamma$ system be inconsistent with $D_s^*$ decay at the
$1\sigma$ level (the corresponding $\Delta M(D_s\gamma)$ must
deviate from the expected value for this decay by more than
$4.4\;$MeV/c$^2$), and that the momentum of the $\pi^0$ be
inconsistent with the $D_{sJ}(2463)\to D_s^*\pi^0$ transition,
also at the $1\sigma$ level. Using these cuts, we see no evidence
for a signal in this mode.

We note that our upper limit for $\Dsjone\to D_s^+\pi^+\pi^-$ is
considerably smaller than the BEH prediction. For this prediction
they calculate both the isospin-violating $D_s^{*+}\pi^o$ rate and
the decay into $D_s^{*+}$ and a virtual $\sigma$ meson that
materializes as a $\pi^+\pi^-$ pair. Although this is a difficult
calculation, we should not be far from seeing this decay.

\section{Conclusions}

CLEO confirms the $c\overline{s}$ state near 2317 MeV discovered by BaBar,
and measures a mass difference with respect to the $D_s^+$ of
350.0$\pm$1.2$\pm$1.0 MeV.  This state is likely to have $J^P=0^+$.

CLEO has made the first observation of a new state near 2463 MeV
and has measured $M({\Dsjone}) - M(D_s^+) = 351.2\pm 1.7\pm 1.0$
MeV. This is likely to be a $1^+$ state. The mass splittings are
consistent with being equal, as predicted by BEH; the difference
[($(1^+-1^-)-(0^+-0^-$)] being 1.2$\pm$2.1 MeV. The two states are
narrow and we limit the total decay widths of both of them to be
$\Gamma<7$ MeV.

We also do not see evidence for any narrow states
in $D_s^{\pm}\pi^{\mp}$ or $D_s^{\pm}\pi^{\pm}$,
which argues against a molecular interpretation.

Theoretical applications of QCD, including exploitation of lattice QCD,
sum rules, and heavy quark and chiral symmetries, are necessary to
extract information on fundamental parameters in the quark sector.  By
coupling HQET with chiral symmetry, the BEH model yielded
predictions about masses, widths and decay modes that were in
conflict with conventional thinking based on potential models.
The experimental results reported here provide powerful support
for the BEH approach.

\section*{Acknowledgments}
Support for this effort was provided by the U. S. National
Science Foundation and the Dept. of Energy.  D. Cinabro,
S. Li, and J. C. Wang contributed greatly to this analysis.
Useful conversations were held with W. Bardeen,
T. Barnes, E. Eichten, C. Hill and J. Rosner.



\label{StoneEnd}


\begin{thebibliography}{99}
\bibitem{noteonxfeed}     The calculation of cross feed backgrounds did
                          not account for the possibility of
                          $\Dsjone\to\Dsjzero \gamma$ decays.  However,
                          based on our direct search for this decay, the
                          impact of ignoring this channel is insignificant.
\end{thebibliography}
\end{document}